\documentclass[sigconf]{acmart}

\usepackage[utf8]{inputenc}
\usepackage{caption}
\usepackage{subcaption}
\usepackage{url}
\usepackage{float}
\usepackage{amsmath}
\usepackage{hyperref}
\usepackage{graphicx}
\usepackage{markdown}
\usepackage{amsthm}
\usepackage{appendix}
\usepackage{xcolor}
\usepackage{enumitem}
\usepackage{tikz}
\usetikzlibrary{arrows}
\usetikzlibrary{patterns}
\usetikzlibrary{positioning}
\usepackage{multirow}
\usepackage{algorithm}
\usepackage{algorithmicx}
\usepackage[noend]{algpseudocode}
\usepackage{appendix}
\usepackage{tikz}
\newtheorem{defn}{Definition}

\newcommand{\signed}[1]{\langle #1 \rangle}
\newcommand{\plocal}{\Pi}

\newcommand{\myparagraph}[1]{\vspace{5pt}\noindent\textbf{#1}}

\algdef{SxnE}[IF]{Upon}{EndUpon}[1]{\textbf{upon event}\ #1\ \algorithmicdo}
\algdef{SxnE}[IF]{Init}{EndInit}{\textbf{Init:}}

\newcommand{\tb}{{\sf TrustBoost} }
\newcommand{\tbnosp}{{\sf TrustBoost}\ignorespaces}
\newcommand{\app}{{\sf App} }
\newcommand{\appx}{{\sf AppX} }
\newcommand{\tblite}{{\sf TrustBoost-Lite} }
\newcommand{\tblitenosp}{{\sf TrustBoost-Lite}\ignorespaces}
\newcommand{\ibc}{CCC}
\newcommand{\propose}{\mathsf{Propose}}
\newcommand{\vote}{\mathsf{Vote}}

\newcommand{\submit}{{\sf submit} }

\newcommand*\circled[1]{\tikz[baseline=(char.base)]{
            \node[shape=circle,draw,inner sep=.1pt] (char) {\textbf{#1}};}}

\definecolor{LightCyan}{rgb}{0.88,1,1}

\copyrightyear{2023} 
\acmYear{2023} 
\setcopyright{acmlicensed}\acmConference[CCS '23]{Proceedings of the 2023 ACM SIGSAC Conference on Computer and Communications Security}{November 26--30, 2023}{Copenhagen, Denmark}
\acmBooktitle{Proceedings of the 2023 ACM SIGSAC Conference on Computer and Communications Security (CCS '23), November 26--30, 2023, Copenhagen, Denmark}
\acmPrice{15.00}
\acmDOI{10.1145/3576915.3623080}
\acmISBN{979-8-4007-0050-7/23/11}

\begin{document}

\title{\tbnosp: Boosting Trust among Interoperable Blockchains}

\author{Peiyao Sheng}
\authornote{The first two authors contributed equally to this work. For correspondence on the
paper, please contact Xuechao Wang at xuechaowang@hkust-gz.edu.cn.}
\email{psheng2@illinois.edu}
\affiliation{%
  \institution{University of Illinois Urbana-Champaign}
  \country{USA}
}

\author{Xuechao Wang}
\authornotemark[1]
\email{xuechaowang@hkust-gz.edu.cn}
\affiliation{%
  \institution{The Hong Kong University of Science and Technology (Guangzhou)}
  \country{China}
}

\author{Sreeram Kannan}
\email{ksreeram@ece.uw.edu}
\affiliation{%
  \institution{University of Washington}
  \country{USA}
}

\author{Kartik Nayak}
\email{kartik@cs.duke.edu}
\affiliation{%
  \institution{Duke University}
  \country{USA}
  }

\author{Pramod Viswanath}
\email{pramodv@princeton.edu}
\affiliation{%
 \institution{Princeton University}
 \country{USA}
 }

\begin{abstract}
Currently there exist many blockchains with weak trust guarantees, limiting applications and participation.  Existing solutions to boost the trust  using a stronger blockchain, e.g., via checkpointing, requires the weaker blockchain to give up sovereignty. In this paper, we propose a family of protocols in which multiple blockchains interact to create a {\em combined} ledger with {\em boosted} trust. We show that  even if several of the interacting blockchains cease to provide  security guarantees, the combined ledger continues to be secure -- our \tb protocols achieve the optimal threshold of tolerating the insecure blockchains. 
This optimality, along with the necessity of blockchain interactions, is formally shown within the classic shared memory model, tackling
the long standing open challenge of solving consensus in the presence of both Byzantine objects and processes.
Furthermore, our proposed construction of \tb simply operates via smart contracts and require no change to the underlying consensus protocols of the participating blockchains, a form of ``consensus on top of consensus''. 
The protocols are lightweight and can be used on specific (e.g., high value) transactions; we  demonstrate the practicality by  implementing and deploying  \tb as cross-chain smart contracts in the {\sf Cosmos} ecosystem using approximately 3,000 lines of {\sf Rust} code, made available as open source \cite{TrustBoostcode}. Our evaluation shows that using 10 {\sf Cosmos} chains in a local testnet, \tb has a gas cost of roughly \$2 with a latency of 2 minutes per request, which is in line with the cost on a high security chain such as {\sf Bitcoin} or {\sf Ethereum}.
\end{abstract}



\maketitle

\section{Introduction}

\noindent {\bf Motivation.}
Currently there exist more than a thousand (layer 1) blockchains, each with its own trust/security level. Blockchains with weak trust guarantees tend to support  limited applications. 
A  common solution for new/weak blockchains is to ``borrow'' trust from a secure chain. A standard way of lending such trust is via checkpointing~\cite{karakostas2021securing,rana2022optimal,sankagiri2021blockchain,tas2022bitcoin} -- here checkpoints attest to the hash of well-embedded blocks every so often and newly mined blocks follow the checkpoints. For instance, {\sf Bitcoin} itself was secured by checkpointing by Nakamoto themselves until as late as 2014.  A critical point to note is that this form of trust lending involves the very consensus layer  of the weak blockchain – the fork choice rule of the weak chains needs to obey the checkpoints. The asymmetrical nature of this approach constrains its applicability, leading to a one-way transfer of trust from stronger to weaker blockchains. Consequently, participants in the weaker chains are often sidelined, losing their ability to influence consensus decisions entirely. In practice, in the {\sf Cosmos} ecosystem newer and application-specific chains (called ``{\sf Cosmos} Zones'') can use the same validator set as the original {\sf Cosmos} chain (called ``{\sf Cosmos} Hub'') via a governance proposal~\cite{interchain} -- in return for the trust of the Hub, the Zones give up their individual sovereignty.


\noindent {\bf Our goal.} This state of affairs begets the following question: how should multiple blockchains interact to create a {\em combined} ledger whose  trust is ``boosted''? Ideally, the ``trust boost'' operations (i.e., deciding which specific transactions or applications need to be in the combined ledger and thus enjoy boosted trust-levels) should be simply offered via smart contract operations without altering the consensus layer (i.e., constituent blockchains do not give up their individual sovereignty while collaboratively contributing to the enhanced consensus). Technically speaking, this means answering the following open question: given $m$ multiple blockchain ledgers, $f$ of which are faulty, i.e., without security guarantees, can we combine them in such a way that  there is consensus on the combined ledger? Note that the adversary can collude across the $f$ faulty blockchains. For simplicity's sake, we initially assume all ledgers share the same security level and will discuss the protocol's generality later on. Answering this question comprehensively, from impossibility results on trust boosting to a concrete protocol with optimal trust boosting properties to a full-stack implementation in the {\sf Cosmos} ecosystem are the goals of this paper. 

\noindent {\bf Blockchain bridges.}
There are two distinct approaches to  boosting trust depending on whether the interaction between the blockchain ledgers is  passive  or  active. In the passive mode, there is no communication between the ledgers and a single combiner has read-access to the ledgers and works to form a combined ledger. 
In the active mode, cross-chain communication (CCC) is allowed across the ledgers via bridges. This approach has only been made possible recently as blockchains have become more interoperable -- recent  CCC projects include  IBC by {\sf Cosmos}~\cite{ibc}, XCM by Polkadot~\cite{xcm}, and CCIP by Chainlink~\cite{ccip}. These bridges allow information to be imported across smart contracts residing on the different programmable blockchains -- the trust combiner we are envisioning is a smart contract too, residing on {\em each} of the blockchains.

\noindent {\bf Main contributions.} 
We examine the multi-chain framework within the classic shared memory model in distributed computing~\cite{herlihy2020art}, where blockchain clients act as processes and blockchain ledgers serve as shared objects. Moreover, in the active mode, we extend the model to enable communication between objects, capturing the functionality of CCC. This extension aligns seamlessly with today's multi-chain frameworks, as programmable and interoperable blockchains currently form a network of interconnected global computers. Our study of blockchain interactions brings to sharp focus the nuanced models in terms of the interactions needed to achieve combined trust. 
As the primary theoretical result, we present a formal proof establishing the necessity of CCC for achieving consensus on top of blockchains. To the best of our knowledge, our work is the first to tackle the long standing open problem of solving consensus in the presence of both Byzantine objects and an infinite number of Byzantine processes in the shared memory model. Our key insight involves utilizing ``powerful" shared objects (i.e., programmable and interoperable blockchains), to address this complex problem.

Specifically, in the passive mode, we show that consensus on  combined ledgers, for any possible combination, is impossible if $f > 0$. Indeed, one of the earlier efforts in the literature~\cite{fitzi2020ledger}, tried to create a combined ledger passively without success.  Focusing on a weaker form of consensus that gives up the total ordering property (while still being able to implement the functionality of a cryptocurrency), referred to as Asynchronous Blockchain without Consensus (ABC)~\cite{sliwinski2019abc},  we show the following in the passive mode.  First, even ABC is impossible, if $m \leq 3f$.  Second, we propose a protocol called \tblite that combines different ledgers to achieve ABC whenever $m>3f$.  
In the active mode, we show that consensus is impossible in a partial synchronous network if $m \leq 3f$. When $m > 3f$, we propose a protocol called \tb that securely combines the $m$ ledgers together. 
Both \tb and \tblite protocols can be viewed as BFT consensus protocols: consensus is now amongst the programmable blockchains (whose actions are executed by smart contracts) communicating over pairwise authenticated channels provided via the CCC infrastructure -- a form of ``consensus on top of consensus''.  

\begin{figure}
    \centering
    \includegraphics[width=0.4\textwidth]{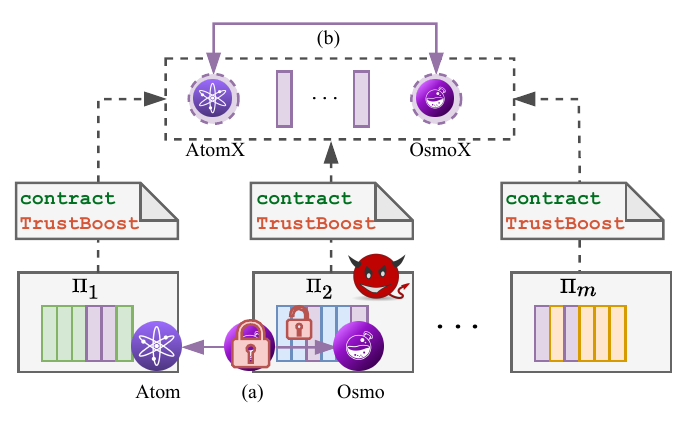}
    \caption{(a) Token exchange across chains are vulnerable to single-chain attacks. Suppose attackers lock 100 Osmos on the {\sf Osmosis} chain in exchange for 10 Atoms on the {\sf Cosmos} chain. Once Atoms are received, a double-spend attack on the transaction which locks 100 Osmos on the {\sf Osmosis} chain leads to 10 ``free'' Atoms, creating a security attack on the {\sf Cosmos} chain. (b) \tb secures contract states. Any application contract (e.g., Atom token contract) can be upgraded to a \tb cross-chain contract (e.g., AtomX) by creating secure global states. The exchange of \tb cross-chain tokens are now  secured by the interacting  blockchains. }
    \label{fig:intro}
\end{figure}

\tb is a lightweight consensus protocol, executable entirely as a smart contract on each of the blockchains. Further, any specific transactions of any application contract can be upgraded using \tb to avoid single-chain security attacks (an example is depicted in Fig.~\ref{fig:intro}). We demonstrate the practicality by implementing and deploying \tb as cross-chain smart contracts in {\sf Cosmos} ecosystem using approximately 3,000 lines of {\sf Rust} code, made available as open source \cite{TrustBoostcode}. 

Several limitations of  smart contract programming impose challenges to implementing BFT consensus protocols using them: (a) contracts only behave passively and we need to ensure that every operation in \tb is properly triggered by some IBC message; (b) contracts work  only with single-threading, preventing parallelism in operations; (c)  {\sf Cosmos}-SDK allows a smart contract to send IBC messages only when a function returns -- a major implementation hurdle, which we deal with by queuing all the IBC messages for each function that need to be sent and send them all when the function returns; (d) finally, special attention should be paid to self-delivered messages. The design principles of our successful  implementation of BFT consensus protocols via smart contracts  might be of independent and broader interest.  

The performance of \tbnosp, particularly latency and gas usage, depends on both the implemented BFT consensus protocols and IBC efficiency. We implement Information Theoretic HotStuff~\cite{abraham2020information} to avoid expensive operations on signature verification, which however leads to an $O(m^2)$ boost in gas usage and a linear increase in latency. Meanwhile, in {\sf Cosmos} a single IBC message (e.g., for cross chain token transfer) would take 2 seconds and cost 350K gas. 
Concretely, with 10 {\sf Cosmos} chains in a local testnet, the total gas cost is roughly \$2 with a latency of 2 minutes when using \tb to boost the security of a standard contract {\sf NameService}\cite{nameservice} -- here gas fees in fiat are extracted from the exchange rate and the gas price of {\sf Osmosis}, a popular  {\sf Cosmos} Zone at the time of writing (April 2023) and are in line with the gas fees of a high security chain such as {\sf Bitcoin} or {\sf Ethereum}. Improving the efficiency of the implemented BFT protocols and IBC would make \tb more performant.



\section{Related Works} 

In this section, we survey related works encompassing blockchain protocols that borrow or boost trust, hierarchical consensus frameworks, distributed computing models, and blockchain interoperability.

\noindent {\bf Borrowing trust.} Checkpointing is a method that allows the trust of a highly secure blockchain to be extended to weaker or newer blockchains~\cite{sankagiri2021blockchain,karakostas2021securing,rana2022optimal}. Validators of a weaker chain periodically submit block hashes and signatures as checkpoints to a more secure chain, and the finality rule of the weaker chain is modified to respect the checkpoints. Consequently, the weaker chain has a slightly slower finality rule - confirming the chain up to the latest checkpoint, which has the same latency and security level as the secure chain. A concrete and practical instantiation of this idea in the context of bringing {\sf Bitcoin} trust to {\sf Cosmos} Zones is \cite{tas2022bitcoin}. A very recent work \cite{tas2023interchain} generalizes checkpointing approaches and offers new insights that align with the design principles underlying our work. The proposed protocol lets a consumer chain draw additional security from a series of provider chains through sequential checkpointing operations. However, this approach necessitates that all chains remain live to guarantee eventual liveness. In contrast, our setting enhances not only the security but also the liveness of all participating chains.

\noindent {\bf Boosting trust.} An early work on robust ledger combining is~\cite{fitzi2020ledger}; parallel ledgers process a common set of transactions independently, and confirmation in the combined ledger is carried out by observers who can read from all ledgers. Similar to \tbnosp, the combine ledger functions even if a certain fraction of underlying ledgers no longer provide any security guarantees. However, the combined ledger only ensures a notion of relative persistence, which is not sufficient even for a payment system, so its practical use is limited. A detailed exploration of this limitation is discussed in Appendix \ref{sec:ledger-combiner}.

\noindent {\bf Hierarchical consensus.} To the best of our knowledge, Steward~\cite{amir2006scaling} was the pioneering work that proposed a concept of ``consensus on top of consensus''. Steward employs a BFT protocol within each local site and a benign fault-tolerant protocol among wide area sites. Each local site, consisting of several potentially malicious replicas, is converted into a single logical trusted participant in the global protocol. GeoBFT~\cite{gupta2020resilientdb} further improves scalability by introducing parallelization of consensus at the local level, and by minimizing communication between sites. However, in comparison to \tbnosp, both Steward and GeoBFT assume a honest supermajority in each local site, which significantly simplified the problem. Furthermore, alterations in the local consensus are needed, whereas in \tbnosp, the global consensus is lightweight and implemented solely through smart contracts.

The idea most closely related to ours is the ``recursive Tendermint''~\cite{retendermint} proposed by the {\sf Cosmos} team, in which Tendermint is run on multiple {\sf Cosmos} chains using the IBC protocol instead of TCP/IP in a peer-to-peer network. However, this concept was only presented as a preliminary idea, without delving into the scientific and engineering challenges that we addressed in \tbnosp.

\noindent {\bf Shared memory vs. message passing.} The shared memory model and message passing model are two fundamental approaches in distributed computing~\cite{herlihy2020art}. In the shared memory model, processes communicate by reading and writing to shared objects, whereas in the message passing model, processes exchange messages with one another to coordinate their actions. Consensus, a critical problem in distributed systems, has also been extensively investigated in both the shared memory model~\cite{herlihy1991wait,gafni2003disk,chockler2002active,abraham2004byzantine} and message passing model~\cite{fischer1985impossibility,fischer1986easy,dwork1988consensus,lamport2001paxos,lamport1982byzantine,lamport1982byzantine,castro1999practical}, yielding a variety of positive and negative results. In this work,  we expand the shared memory model to enable communication between objects, an adaptation that aligns seamlessly with today’s multi-chain framework. Furthermore, we establish theoretical bounds in this refined model and present a clear delineation of the interactions required to achieve combined trust among blockchains.

\noindent {\bf Blockchain interoperability.} \cite{zamyatin2021sok} presents a general framework to design and evaluate CCC protocols that facilitate blockchain interoperability. The most significant application of blockchain interoperability is atomic cross-chain swaps~\cite{herlihy2018atomic,thyagarajan2022universal}, which enable the exchange of assets across multiple distinct blockchains. However, these protocols necessitate intricate and time-consuming user interactions with the blockchains and their peer-to-peer transaction nature often results in lower liquidity compared to centralized exchanges.

In order to facilitate general cross-chain applications, cross-chain bridges have emerged as a significant building block in today's multi-chain world. There are three primary categories: 1) committee-based bridges; 2) optimistic bridges; 3) light client bridges. Committee-based bridges (PolyNetwork~\cite{ploynetwork}, Wormhole~\cite{wormhole}, LayerZero~\cite{layerzero}, CCIP~\cite{ccip}, etc.) employ a trusted committee of validators to sign off on state transfers, with security relying on the honest majority assumption. Optimistic bridges (like Nomad~\cite{nomad} and Near’s Rainbow Bridge~\cite{rainbow}) require participants to deposit collateral, and depend on a watchdog service to continuously monitor the blockchain and confiscate offenders’ collateral upon detecting invalid updates. However, optimistic protocols fundamentally demand long confirmation latency to ensure high probability of detecting invalid updates. Light client bridges (e.g., Cosmos IBC~\cite{ibc}) are trustless, using on-chain light clients to verify state transitions on the other blockchain. Zk-SNARKs are further leveraged to enhance the efficiency of state verification~\cite{xie2022zkbridge,snark}. While \tb can utilize all these bridge types, light-client bridges are preferred due to their trustlessness and efficiency. Further improvements in the security and performance of bridges represent an interesting and active research area, but it falls outside the scope of this paper.

A very recent work \cite{xue2022cross} proposes a cross-chain state machine replication protocol in the passive mode, which maintains a consistent state across multiple chains; indeed the security guarantees in \cite{xue2022cross} hold only when {\em each} of the involved blockchains is secure (as expected by  one of our theoretical results (cf.\ Theorem~\ref{thm:passive})). 




\section{Preliminaries}

\subsection{System model}
Our model is inspired by today’s public blockchains, which serve as global state machines shared by blockchain clients. Following the notations of shared memory models in the distributed computing literature~\cite{herlihy2020art}, we consider a shared memory system consisting of a (possibly infinite) collection of \textit{processes} $P_1,P_2,\cdots$ interacting with a finite collection of $m$ \textit{objects} $O_1,O_2,\cdots,O_m$. Both objects and processes are modeled as I/O automata~\cite{lynch1988introduction}. 

A process automaton models a blockchain client such as a full node or light client. Since blockchain clients are untrusted, we assume no trust from the processes. 
In other words, we allow any number of Byzantine processes, but only honest processes enjoy the guarantees of the protocol. An object automaton models a blockchain ledger (or a smart contract), providing three basic interfaces \textit{submit}, \textit{check} and \textit{read}. A process can access the ledger states via the \textit{read} operation and submit transactions to change the states via the \textit{submit} operation. Once a submitted transaction $tx$ is committed, the \textit{check} interface returns true for $tx$.
We assume that at most $f$ out of $m$ objects are Byzantine, potentially providing arbitrary responses. Byzantine objects represent faulty ledgers run by corrupted nodes, whereas ledgers operated by nodes with an honest (super)majority are modeled as correct or honest objects.

Depending on whether the blockchains in the system are interoperable or not, we consider two different mode: an \textit{active mode} and a \textit{passive mode}. In the active modes, objects are fully connected via authenticated channels and exchange messages through these channels. The channels provide two interfaces, {\it send} and {\it deliver}, and they ensure message integrity and reliability, that a message is delivered from $p$ to $q$ if and only if it was previously sent from $p$ to $q$. We assume this fully connected object network is partially synchronous, meaning there is a global stabilization time ($GST$) chosen by the adversary, unknown to honest nodes and also to the protocol designer, such that after $GST$, all messages sent between honest objects are delivered within $\Delta$ time. Before $GST$, the adversary can delay messages arbitrarily.  When $GST=0$, the network becomes synchronous. In practice, authenticated channels are instantiated by cross-chain bridges. In the passive mode, objects do not communicate with each other, and processes have read-only access to objects. 
Note that granting write access to the processes would effectively transform the passive mode into the active mode, as processes can facilitate message forwarding among objects.

Additionally, to model blockchains capable of producing commitment certificates as proofs of confirmation (for example, the quorum certificate in many BFT-SMR protocols), we further assume the existence of a public-key infrastructure (PKI) for the set of objects. It is important to note that the PKI may or may not be required in our construction.

\noindent {\bf Contrast with classic shared memory model.} In the traditional shared memory model, communication between processes is facilitated through a collection of shared objects, such as registers. These objects store values and offer two basic operations: {\it read} and {\it write}. The read operation retrieves the requested values, while the write operation updates the stored values. In our model, an object is extended to store states in a blockchain ledger instead of a single value. Hence, we adapt the interfaces to allow processes to {\it submit} transactions (analogous to write values) and {\it read} states (analogous to read values). The additional interface {\it check} returns the transaction status, indicating whether a {\it submit} operation has been successfully completed. Our approach aligns with the regular register model, where a read operation can return the value written by either the most recently concluded write operation or a write operation with which it overlaps~\cite{lamport1985interprocess}. Our passive mode represents a simplified version of the classic shared memory model, with processes having read-only access to objects. This mirrors the ``lazy'' trust-boosting approach, where each blockchain client has read-access to the ledgers and works to form a combined ledger. On the other hand, our active mode is more nuanced and presents a clear distinction from the classic model. In the traditional shared memory model, objects are passive, meaning they undergo state changes only in response to process requests. However, in our active mode, objects can communicate with other objects or observe their states. 
This advanced shared object capability has been made possible by recent developments in blockchain interoperability. It also improves the system's modularity, as the message-passing operations between objects encapsulate a pair of read and write operations between different objects and processes. Additionally, our active mode can also be considered as a new hybrid model that combines shared memory across processes and message passing across objects, whereas in classic distributed computing, only messages among processes are considered.

\subsection{Consensus problems}

To study the security guarantee of the ``meta'' consensus protocol built on top of blockchains, we provide definitions for several relevant problems, ranging from binary consensus to ledger consensus, including their relaxed versions.

\noindent {\bf Binary consensus.} We start with the problem of \textit{binary consensus}, 
where each object starts with some initial value (0 or 1) and all (honest) processes try to commit the same value by the end of the protocol. 
\begin{defn}[Binary Consensus]
\label{def:consensus}
\hfill
\begin{itemize}
    \item \textbf{Agreement}: No conflicting values are committed by honest processes.
    \item \textbf{Validity}: If every honest object starts with the same value, this value will be committed by honest processes.
    \item \textbf{Termination}: Every honest process commits one of the values. 
\end{itemize}
\end{defn}

We are also interested in a relaxed notion of binary consensus, called \textit{ABC binary consensus}, derived from~\cite{sliwinski2019abc}. The key insight leading to this relaxation is that in numerous applications, termination is only required in optimistic scenarios. 

\begin{defn}[ABC Binary Consensus]
\label{def:abc}
\hfill
\begin{itemize}
    \item \textbf{Agreement}: Same as in Definition~\ref{def:consensus}. 
    \item \textbf{Validity}: Same as in Definition~\ref{def:consensus}. 
    \item \textbf{Honest termination}: If every honest object starts with the same value, this value will be committed by honest processes. 
\end{itemize}
\end{defn}


\noindent {\bf Ledger consensus.} 
It is known that binary consensus protocols can be used to solve the problem of \textit{ledger consensus}, i.e., \textit{state machine replication (SMR)}~\cite{smr}, where all processes maintain a list of transactions that grows in length, called a \textit{public ledger} (cf. Definition~\ref{def:pl}), with the help of the shared objects. Each process initially starts with the same state and updates the state by executing all transactions in its ledger. 
We first define the notion of a public ledger.
\begin{defn}[Public ledger]
\label{def:pl}
A public ledger $\mathcal{L}$ is a growing list of transactions that provides the following two process interfaces:
\begin{itemize}
    \item Submit: a process can submit a transaction $tx$ to $\mathcal{L}$ by calling {\sf submit}$(tx)$.
    \item Check: a process can check whether $tx \in \mathcal{L}$ by calling {\sf check}$(tx)$. If {\sf check}$(tx)$ returns true, the process will commit $tx$.
    \item Read: a process can access the set of states $S$ stored on $\mathcal{L}$ by calling {\sf read}$(S)$.
\end{itemize}

\end{defn}

In the problem of ledger consensus, we allow the following synchronous out-of-band communications among processes: 1) processes can send transactions to objects, but still processes send no other message to objects; 2) processes can communicate with each other only to prove the confirmation of certain transactions (i.e., allowed to sync up state with each other or bootstrap new processes).
The guarantees of ledger consensus are defined as follows.

\begin{defn}[Ledger Consensus]
\label{def:ledger}
\hfill
\begin{itemize}
    \item \textbf{Agreement}: If some honest process commits $tx$, every honest process will also commit $tx$; Moreover, $tx$ appears at the same place in the ledgers of all honest processes. Equivalently, if $[tx_0, tx_1, \cdots, tx_i]$ and $[tx_0', tx_1', \cdots, tx_{i'}']$ are two ledgers output by two honest processes, then $tx_j = tx_j'$ for all $j \leq \min(i,i')$.
    \item \textbf{Termination}: If a process submits $tx$ to all honest objects, $tx$ will be committed by all honest processes. 
\end{itemize}
\end{defn}

Similarly, ABC can also be used to build a public ledger, although without total ordering of the transactions. However, this suffices to implement the functionality of a cryptocurrency like {\sf Bitcoin} as shown in~\cite{sliwinski2019abc,baudet2020fastpay,guerraoui2019consensus}. Suppose the transaction space is now equipped with a conflict relation (e.g. double spend transactions in {\sf Bitcoin}), then the problem of ABC ledger consensus is defined as follows.

\begin{defn}[ABC Ledger Consensus]
\label{def:abc_ledger}
\hfill
\begin{itemize}
    \item \textbf{Weak agreement}: No conflicting transactions are committed by honest processes.
    \item \textbf{Honest termination}: If a process submits $tx$ to all honest objects and there are no conflicting transactions, $tx$ will be committed by all honest processes.
\end{itemize}
\end{defn}


\noindent {\bf Contrast with consensus in traditional models.} As mentioned earlier, our passive mode is a simplified version of the classic shared memory model. Surprisingly, as we will demonstrate in \S\ref{sec:converse}, ABC remains achievable even within this highly restricted model. This observation leads to the development of the protocol \tblite in \S\ref{sec:lite}, where no CCC is needed. As for our active mode, it is better to compare it with the classic message passing model, where $m$ validators communicate to reach consensus. It is not difficult to see that any consensus protocol $\Pi$ in the message passing model can be used to solve consensus in the active mode: Let $3f+1$ objects implement $\Pi$; each process commits a value/transaction if it is read from at least $2f+1$ objects (note that solving consensus requires $m>3f$ in partial synchrony). This reduction allows us to construct \tb in a black-box manner. Conversely, solving consensus in our active mode does not necessarily solve consensus in the classic message passing model. Indeed, in the above example, there are $m-3f-1$ objects unused when solving consensus in our active mode, but all $m$ validators need to participate and reach consensus in the message passing model.

In summary, our active mode offers a more nuanced and flexible approach, combining aspects of shared memory and message passing models, and leveraging advancements in blockchain interoperability. While it can utilize traditional consensus protocols, it can also achieve consensus with fewer objects and additional efficiency, making it a distinct and innovative approach compared to traditional models.

\section{Theoretical Bounds for the Shared Memory Model}
\label{sec:converse}

In a seminal work~\cite{loui1987memory}, an impossibility result for the binary consensus problem is shown for an asynchronous shared memory system, where a single process may crash. In this work, we circumvent this impossibility result by (1) adopting a weaker notion of consensus in our passive mode, and (2) permitting partial synchronous communication among objects in the active mode.

Previous works~\cite{lamport2001paxos,chockler2002active,abraham2004byzantine} circumvent this impossibility result by employing a leader oracle. The state-of-the-art study~\cite{abraham2004byzantine} proves that $1/3$ is the optimal resilience of Byzantine objects for solving consensus; however, it only takes benign processes into account. A recent work~\cite{cohen2021tame} considers Byzantine processes but only focus on problems weaker than consensus, such as reliable broadcast, snapshots, and asset transfer (essentially equivalent to ABC). To the best of our knowledge, our work stands as the first attempt to tackle the long standing open problem of solving consensus in the presence of both Byzantine objects and an infinite number of Byzantine processes in the shared memory model.

Specifically, in this section, we show that in the passive mode (i.e., without communication among objects), binary consensus (cf. Definition~\ref{def:consensus}) is impossible, while ABC binary consensus (cf. Definition~\ref{def:abc}) can be still achieved. Meanwhile, in the active mode, both problems can be solved as long as $m>3f$. We adapt proof techniques from classic distributed system problems~\cite{dwork1988consensus,fischer1985impossibility} to our nuanced model and show the following tight results and sharp distinctions between consensus and ABC. 

\subsection{Passive mode}

We start with the theoretical bounds in the passive mode.

\begin{theorem}
\label{thm:passive}
In the passive mode, we have 
\begin{enumerate}
    \item Consensus can be achieved if and only if $f=0$.
    \item ABC can be achieved if and only if $m>3f$.
\end{enumerate}
\end{theorem}





\begin{proof}
Since in our model, the consensus protocols may or may not use PKI, we prove the strongest results: for the negative results (i.e., ``only if'' parts), we assume PKI is used; and for the positive results (i.e., ``if'' parts), we avoid using PKI.

\hfill

\noindent (1) Consensus:

\noindent \textit{Only if}: 
Seeking a contradiction, let us assume there is a protocol that solves consensus in the passive mode. Divide all the processes into two sets: $X$ and $Y$, each with at least one honest process. We first construct the following $m+1$ worlds. See Fig.~\ref{fig:converse2}.

\begin{figure*}
    \centering
    \includegraphics[width=0.8\textwidth]{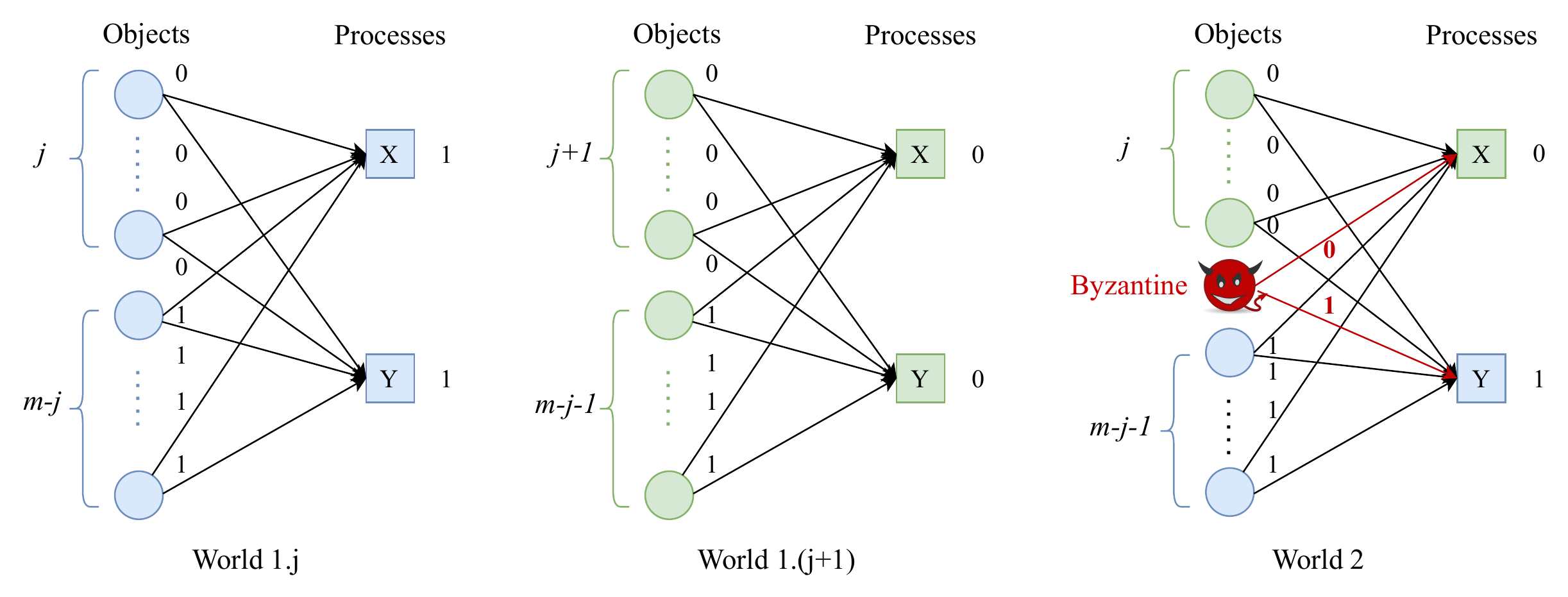}
    \caption{Different worlds in the proof of Theorem~\ref{thm:passive}. 
    }
    \label{fig:converse2}
\end{figure*}

\noindent {\bf World $1.i$ ($0\leq i \leq m$):} In world $1.i$, the first $i$ objects start with value 0 and the rest $m-i$ objects start with value 1. By termination, in all $m+1$ worlds, processes in $X$ and $Y$ must eventually commit some value. By validity, the committed value must be 1 in world $1.0$ and 0 in world $1.m$. Then there must exist some integer $0\leq j\leq m-1$ such that the committed value is 1 in world $1.j$ and 0 in world $1.(j+1)$. 
Now consider the following world:

\noindent {\bf World $2$:} World $2$ will be a hybrid world where the view of processes in $X$ in this world will be indistinguishable to their views in world $1.(j+1)$ and the view of processes in $Y$ in this world will be indistinguishable to their view in world $1.j$. In world $2$, the first $j$ objects start with value 0 and the last $m-j-1$ objects start with value 1. The adversary will use its Byzantine power to corrupt the $(j+1)$-th object to perform a split-brain attack and make $X$ and $Y$ each believe that they are in their respective worlds. The $(j+1)$-th object will equivocate and act as if its starting value is 0 when communicating with $X$ and as if its 1 when communicating with $Y$.
And since there is no communication among the objects, their views will be indistinguishable to the views in worlds $1.j$ and $1.(j+1)$. Moreover, the view of $X$ in this world will be indistinguishable to the view of $X$ in world $1.(j+1)$ and the view of $Y$ in this world will be indistinguishable to the view of $Y$ to world $1.j$. Therefore, $X$ will commit 0 and $Y$ will commit 1. This violates the agreement property.

\hfill

\noindent \textit{If}: With $f=0$, we solve the problem of consensus in the passive mode. The protocol is simple: each process queries the value from all $m$ objects; and the process commits a value if it receives the same value from all $m$ objects; otherwise if the process receives both 0 and 1, it commits 0 by default. It is easy to check that this protocol satisfies agreement, validity and termination; Moreover, there is no communication among the objects.

\hfill

\noindent (2) ABC:

\noindent \textit{Only if}: Seeking a contradiction, let us assume there is a protocol that claims to solve ABC with $f\geq m/3$ Byzantine objects. Divide the $m$ objects into three sets: $A$, $B$, and $C$, each with at least one object and at most $f$ objects. Divide all the processes into two sets: $X$ and $Y$, each with at least one process. We consider the following three worlds and explain the worlds from the view of $A$, $B$, $C$, $X$ and $Y$.

\noindent {\bf World 1:} In World 1, objects in $A$ and $B$ start with the value 1. Objects in $C$ are Byzantine but pretend to be honest with initial value 0. Since $C$ has at most $f$ objects, the processes in $X$ must eventually commit a value by honest termination. For validity to hold, all the processes in $X$ will output 1. 

\noindent{\bf World 2:} In World 2, objects in $B$ and $C$ start with the value 0. Objects in $A$ are Byzantine but pretend to be honest with initial value 1. Since $A$ has at most $f$ objects, the processes in $Y$ must eventually commit a value by honest termination. For validity to hold, all the processes in $Y$ will output 1.

\noindent{\bf World 3:} World 3 will be a hybrid world where the view of $X$ in this world will be indistinguishable to the view of $X$ in world 1 and the view of $Y$ in this world will be indistinguishable to the view of $Y$ in world 2. $A$ will start with value 1 and $C$ will start with value 0. The adversary will use its Byzantine power to corrupt $B$ to perform a split-brain attack and make $X$ and $Y$ each believe that they are in their respective worlds. $B$ will equivocate but act honestly as if its starting value is 1 when communicating with $X$ and as if its 0 when communicating with $Y$. Then by an indistinguishability argument, $X$ will commit 1 and $Y$ will commit 0. This violates the agreement property.

\hfill

\noindent \textit{If}: Assume $m>3f$, now we solve the problem of ABC in the passive mode. The protocol is simple: each process queries the value from all $m$ objects; and the process commits a value if it receives the same value from at least $2f+1$ objects. It is easy to check that this protocol satisfies agreement, validity and honest termination; Moreover, there is no communication among the objects.
\end{proof}


\subsection{Active mode}

\noindent {\bf Active mode}. For the completeness of our results, we prove the following theorem in the active mode, which is similar to the well-known impossibility result in the partially synchronous network  model~\cite{dwork1988consensus}. 

\begin{theorem}
\label{thm:active}
In the active mode, we have: 
\begin{enumerate}
    \item Consensus can be achieved if and only if $m>3f$.
    \item ABC can be achieved if and only if $m>3f$.
\end{enumerate}
\end{theorem}

\begin{proof}
Since in our model, the consensus protocols may or may not use PKI, we prove the strongest results: for the negative results (i.e., ``only if'' parts), we assume PKI is used; and for the positive results (i.e., ``if'' parts), we avoid using PKI.
\hfill

\noindent (1) Consensus:

\noindent \textit{Only if}: 
Seeking a contradiction, let us assume there is a protocol that claims to solve consensus with $f\geq m/3$ Byzantine objects. Divide the $m$ objects into three sets: $A$, $B$, and $C$, each with at least one object and at most $f$ objects. Divide all the processes into two sets: $X$ and $Y$, each with at least one process. We consider the following three worlds and explain the worlds from the view of $A$, $B$, $C$, $X$ and $Y$. In all three worlds, we will assume that all messages between $A \longleftrightarrow B$ and $B\longleftrightarrow C$ arrive immediately. See Fig.~\ref{fig:converse1}.

\begin{figure*}
    \centering
    \includegraphics[width=0.8\textwidth]{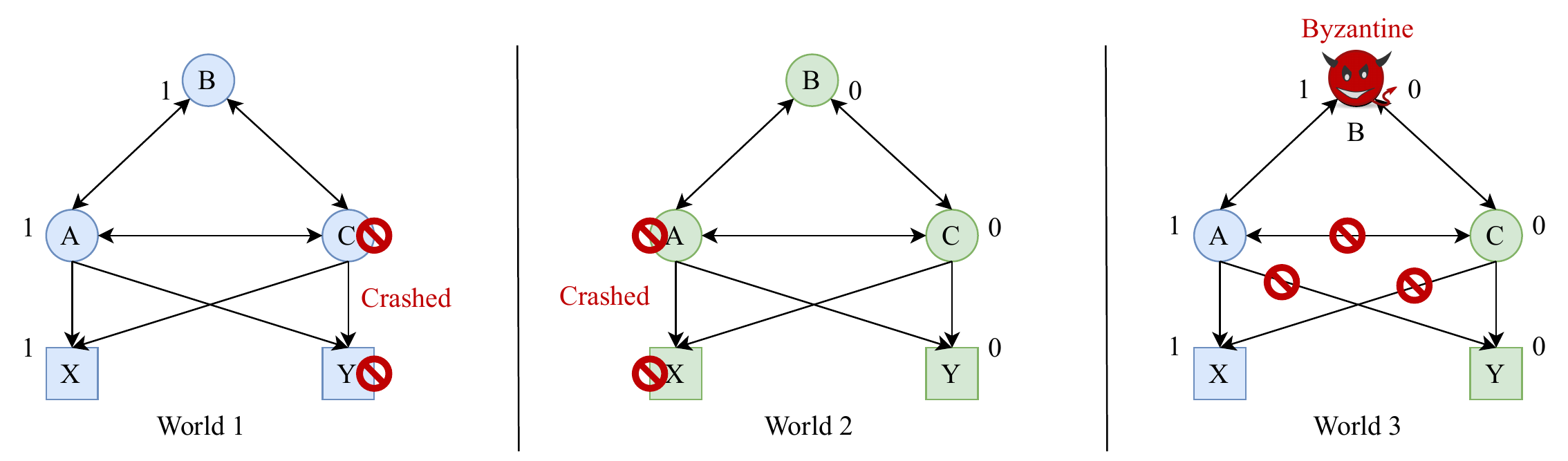}
    \caption{Different worlds in the proof of Theorem~\ref{thm:active}. 
    }
    \label{fig:converse1}
\end{figure*}

\noindent {\bf World 1:} In World 1, objects in $A$ and $B$ start with the value 1. Objects in $C$ and processes in $Y$ have crashed. Since $C$ has at most $f$ objects, the processes in $X$ must eventually commit a value by termination. For validity to hold, all the processes in $X$ will output 1. From the perspective of $A$, $B$ and $X$, they cannot distinguish between a crashed (or Byzantine) $C$ vs. an honest $C$ whose messages are delayed.

\noindent{\bf World 2:} World 2 will be a world similar to world 1 where the roles of $A$ and $C$ and the roles of $X$ and $Y$ are interchanged. The objects in $B$ and $C$ start with the value 0. Objects in $A$ and processes in $X$ have crashed. So, all the processes in $Y$ will output 0 by termination and validity. Again, from the perspective of $B$, $C$ and $Y$, they cannot distinguish between a crashed $A$ vs. an honest $A$ whose messages are delayed. 

\noindent{\bf World 3:} World 3 will be a hybrid world where the view of $A$ and $X$ in this world will be indistinguishable to the view of $A$ and $X$ in world 1 and the view of $C$ and $Y$ in this world will be indistinguishable to the view of $C$ and $Y$ in world 2. $A$ will start with value 1 and $C$ will start with value 0. The adversary will use its Byzantine power to corrupt $B$ to perform a split-brain attack and make $A$ (or $X$) and $C$ (or $Y$) each believe that they are in their respective worlds. $B$ will equivocate and act as if its starting value is 1 when communicating with $A$ and $X$ and as if its 0 when communicating with $C$ and $Y$. If the adversary delays messages between $A \longleftrightarrow C$, $A \longrightarrow Y$ and $C \longrightarrow X$ for longer than the time it takes for $X$ and $Y$ to decide in their respective worlds, then by an indistinguishability argument, $X$ will commit to 1 and $Y$ will commit to 0. This violates the agreement property.

\hfill

\noindent \textit{If}: Assume $m>3f$, now we solve the problem of consensus in active mode. The objects run any partial synchronous consensus protocol (in the classic message passing model) and send the committed value to the clients. The client commits a value if it receives the same value from at least $2f+1$ objects. 
It is easy to check that this protocol satisfies agreement, validity and termination.

\hfill

\noindent (2) ABC: 

\noindent \textit{Only if}: Same as the proof for consensus as above. Note that in world 1\&2 honest objects start with the same value, hence honest termination suffices for the proof.

\hfill

\noindent \textit{If}: Same as the algorithm that solves ABC in the passive mode (in the proof of Theorem~\ref{thm:passive}).
\end{proof}

\noindent {\bf Consensus with reduced communication and connectivity.} So far, we have observed that ABC (cf. Definition~\ref{def:abc}) can be solved without any communication among the objects, while consensus (cf. Definition~\ref{def:consensus}) cannot. One natural question arises: Can we design new consensus protocols to make \tb more efficient in terms of number of IBC connections and messages? Theoretically, it would also be interesting to study the lower bounds on number of messages and network connectivity for consensus in our active mode. For instance, it is evident that when solving consensus with $m>3f+1$ objects, we only need $3f+1$ objects. The extra $m-3f-1$ objects do not even to establish connections with other objects. Meanwhile, in the classic message passing model, all honest objects must form a connected component in order to solve consensus. However, identifying the minimum requirement on network connectivity for consensus in the active mode remains an interesting and challenging problem.

\section{The \tb Protocol}
\label{sec:protocol}

\begin{figure}
    \centering
    \includegraphics[width=0.4\textwidth]{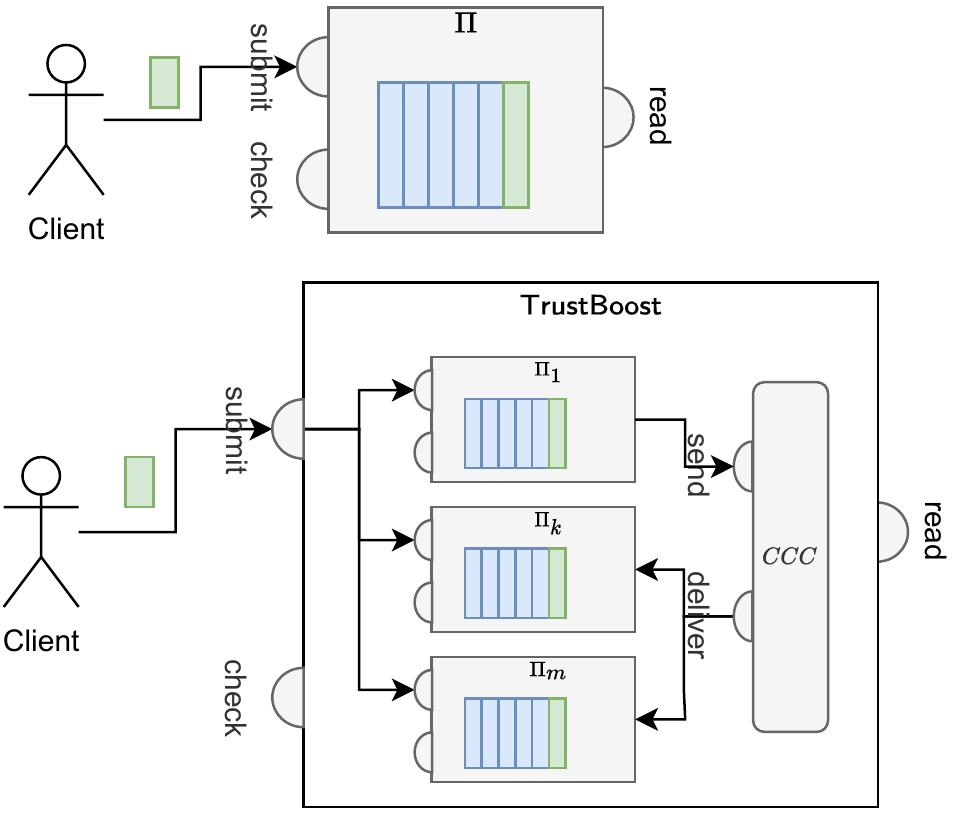}
    \caption{Clients see the same interface of submitting a transaction to \tb as submitting a transaction to a single blockchain.} 
    \label{fig:comparison}
\end{figure}

\subsection{\tbnosp: blockchains as objects}


\tb is run by $m$ blockchains, which support smart contracts, simulating a group of objects written by processes (blockchain clients) to maintain global states. The $i$-th blockchain ($1\leq i \leq m$) is run by a group of nodes with $\beta_i$ fraction of which are Byzantine. Blockchains can be permissioned or permissionless.
Communications between blockchains are made possible by cross-chain communication (\ibc), which is a protocol that manages bidirectional ledger-to-ledger links. There are two primitives used by the \tb protocol. 
\begin{itemize}
    \item \textbf{Local consensus protocol.} $\plocal_k$ is a public verifiable ledger consensus protocol (cf. Definition~\ref{def:ledger}), it provides an interface $\plocal_k.${\sf submit}$(tx)$ (analogous to the write operation of objects), allowing clients to submit and commit transactions to the blockchain. 
    Clients can verify if a transaction $tx$ has been committed to the ledger through $\plocal_k.${\sf check}$(tx)$ and read the current ledger state $S$ through $\plocal_k.${\sf read}$(S)$, which is analogous to the read operation of objects. Optionally, $\plocal_k.${\sf check}$(tx)$ may return a commitment certificate, such as a quorum certificate in many BFT-SMR protocols, as proof of commitment. $\plocal_k$ can be instantiated by a black-box partial synchronous consensus protocol.
    \item \textbf{Cross-chain communication (\ibc).} \ibc\ is a routing protocol that enables independent blockchains to actively communicate with each other. It consists of two primitives, $CCC.${\sf send}$(src, dst, tx)$ and $CCC.${\sf deliver}$(src, dst, tx)$, which transmit transaction $tx$, with $src$ and $dst$ representing the source and destination blockchain identifiers, respectively. \ibc\ ensures both reliability and authenticity. Reliability guarantees that any transaction sent by the source chain will be delivered to the destination chain eventually, while authenticity ensures that any node on the destination chain can verify that the transactions delivered by the protocol correspond to committed states on the source chain.
\end{itemize}


\begin{algorithm}[H]
\caption{Protocol \tb$(CCC, k, \plocal_k)$ }
\label{alg:trustboost}
\begin{algorithmic}[1]
    \Init
        \State $txVotes \gets$ an empty map
        \State $m \gets$ number of participating chains
    \EndInit
    
\Function{{\sf submit}}{$tx$}
\label{func:submit}
    \For{$dst = 1, \cdots, m$}
        \State invoke $\ibc$.{\sf send}$(k, dst, \signed{\propose, tx})$
    \EndFor
\EndFunction

    \Upon{$\ibc$.{\sf deliver}$(src, dst, \signed{\propose, tx})$}\label{line:deliver}
        \State $txVotes[tx] = \emptyset$
        \For{$dst = 1, \cdots, m$}
            \State invoke $\ibc$.{\sf send}$(k, dst, \signed{\vote, tx})$
        \EndFor
    \EndUpon
    \Upon{$\ibc$.{\sf deliver}$(src, k, \signed{\vote, tx})$}
        \State $txVotes[tx] = txVotes[tx] \cup \{src\}$
        \If{$|txVotes[tx]| = \lfloor 2m/3\rfloor + 1$}
            \State invoke $\plocal_k.\mathsf{submit}(tx)$\label{line:submit}
        \EndIf
    \EndUpon
    
    \Function{{\sf check}}{$tx$}\label{func:check}
    \State $cnt\gets 0$
    \For{$k = 1, \cdots, m$} \Comment{These $m$ queries are executed concurrently }
        \If{$\plocal_k.${\sf check}$(tx)$ returns true}
            \State $cnt \gets cnt + 1$
        \EndIf
    \EndFor
    \If{$cnt\ge \lfloor m/3\rfloor + 1$} 
        \State return true
    \EndIf
    \State return false 
\EndFunction

\Function{{\sf read}}{$S$}
\label{func:read}
    \State $tx\gets$ the latest transaction which has $check(tx) = true$ 
    \State $C\gets $ the set of chains whose latest committed transaction is $tx$
    \For{$k \in C$}
        \State $V_k = \plocal_k.{\sf read}(S)$
        \State  return $V_k$ if majority of $C$ agree on $V_k$
    \EndFor
\EndFunction 
        
\end{algorithmic}
\end{algorithm}

\myparagraph{\tb protocol.} With the above primitives, the \tb protocol can be built up by running a public verifiable ledger consensus protocol to issue global states on $m$ local blockchains. It also provides an interface \tbnosp.{\sf submit}$(tx)$ to accept requests from clients, an interface \tbnosp.{\sf check}$(tx)$ for clients to check the commitment of transactions (see Fig.~\ref{fig:comparison}) and an interface \tbnosp.{\sf read}$(S)$ to fetch the current ledger states $S$. By using \ibc\ as transmission channels, and invoking $\plocal_k (1\le k \le m)$ to commit transactions, any partial synchronous consensus protocol can be used to instantiate $\tbnosp$. While local consensus protocols can be either permissioned or permissionless, \tb protocol is a permissioned BFT protocol.

For simplicity, we demonstrate how to instantiate \tb using a majority voting protocol in Algorithm~\ref{alg:trustboost}. Although not a complete consensus protocol, it contains essential building blocks (propose, vote, and commit phases) of most consensus protocols and the usage of all primitives. Initially, the protocol sets up data structures to store votes for transactions. When a client wants to post a transaction $tx$ on \tb blockchains, it calls the \tbnosp.{\sf submit} function (line~\ref{func:submit}), which prompts \ibc\ to broadcast a proposal to all blockchains. Lines~\ref{line:deliver}-\ref{line:submit} detail how to handle cross-chain transactions such as {\sf Propose} and {\sf Vote}. Once the commitment condition is met, for example, in this case, a supermajority of votes are collected, the \tb protocol commits the transaction by requesting the local blockchain to commit the transaction (line~\ref{line:submit}). The protocol also offers a {\sf check} function (line~\ref{func:check}) to derive the global states of a transaction from local states. In this example, the transaction is considered committed by \tb once it is committed by more than $2/3$ of the local blockchains.

More complex rules can be designed according to the specifications of various consensus protocols. For example, when instantiating \tb with BFT protocols that use signatures, we require $\Pi_k$ to provide commitment certificates to enable authentication after being forwarded to a third chain. Conversely, local blockchains can also be permissionless (like public proof-of-stake blockchains) if the consensus protocol used for \tb does not rely on a PKI.

\noindent {\bf Security guarantee.} In \tbnosp, each blockchain operates similarly to a single object, with read and write operations used to access and apply changes to stored states. A blockchain that adheres to the security assumption ($\beta_i < \theta_i$, where $\theta_i$ is the security threshold of $\Pi_i$) is considered an honest blockchain and behaves like an honest object (e.g., no equivocation). Otherwise, it may behave arbitrarily, akin to a Byzantine object, when the security assumption is not met.

Let $f = |\{i: \beta_i \ge \theta_i, 1\leq i \leq m\}|$ represent the number of faulty blockchains. We aim to prove that \tb securely constructs a combined ledger with total ordering as long as $f < t$, where $t$ is the security threshold of \tb.

To formally demonstrate that \tb solves ledger consensus, we must establish that it satisfies the two properties defined in Definition~\ref{def:ledger}.

First, agreement necessitates that if an honest client determines a transaction $tx$ has been committed on \tb, every honest client will also commit it. Since \tb employs a publicly verifiable ledger consensus protocol, when $f < t$, it ensures agreement as long as the underlying blockchains can simulate an honest object. In the SMR literature, an honest object is consistently defined as an object that abides by the protocol's rules. To better understand the descriptions of an honest object, we can think of it as a server that reads inputs from a communication channel and performs actions (such as updating stored states or broadcasting messages through communication channels) according to the protocol specifications. In this context, an honest blockchain with accessible CCC can also read inputs or broadcast messages by invoking CCC's ${\sf deliver}, {\sf send}$ interface and executing actions through smart contracts to apply changes to its states. Moreover, since an honest blockchain runs a publicly verifiable ledger consensus, it guarantees that the blockchain will never reach two conflicting states concerning \tb execution (agreement).

To prove termination, if a client submits a transaction $tx$ to all honest blockchains, the \tb protocol ensures termination, which means it will prompt all honest blockchains to commit $tx$. Subsequently, each honest blockchain will process the $tx$ execution through a local transaction, which will terminate due to the termination property of the local consensus protocol. As a result, the \tbnosp.{\sf check}$(tx)$ will eventually return true.

\subsection{\tblite}
\label{sec:lite}

Here we propose a lightweight protocol called \tblite that solves the problem of ABC ledger consensus as defined in Definition~\ref{def:abc_ledger}. In \tblitenosp, $m$ blockchains are independently run by each group $C_i$ using local consensus protocols $\plocal_i$ ($1\le i \le m$). The transactions in \tblite use the unspent transaction output (UTXO) model, where a transaction consists of a set of inputs and outputs and can be denoted as $tx = (in, out)$. The inputs are pointers to some outputs in previously committed transactions, we use {\it input.from} to represent the transaction that contains the output which the {\it input} points to. \tblite only provides two interfaces, {\sf submit} and {\sf check}, which are defined in Algorithm~\ref{alg:tblite}. The {\sf read} interface is not available in \tblite because there is no consensus on the ledger states.

\myparagraph{\tblite protocol.}  Different from \tbnosp, no cross-chain communication is needed in \tblitenosp. Thus, the requests from clients will trigger the \submit of each local blockchain directly (line~\ref{lite:submit}). And to check whether a transaction is committed by \tblitenosp, the client will observe the states of $m$ individual blockchains and make sure (1) the transaction is committed on at least $2/3$ of local blockchains and (2) all its inputs are outputs from globally committed transactions (line~\ref{lite:check}).

\noindent {\bf Security guarantee.} 
Let $f = |\{i: \beta_i \ge \theta_i, 1\leq i \leq m\}|$ represent the number of faulty blockchains. We aim to prove that \tblite solves ABC ledger consensus. To demonstrate this, we must show that it satisfies the two properties defined in Definition~\ref{def:abc_ledger}.

To prove weak agreement, we assume that an honest client commits $tx$, which means the \textsf{check} function of \tblite returns true when the honest client called it. In other words, at least $ \lfloor 2m/3\rfloor + 1$ of the blockchains have committed $tx$, and there does not exist a committed double-spending input. Among these chains, at least $ \lfloor 2m/3\rfloor + 1 - f$ of the blockchains are honest, and they will commit $tx$ and terminate within a bounded time. Since honest blockchains will never commit a conflicting transaction (agreement), faulty blockchains, together with the remaining honest blockchains, are not enough to commit a conflicting transaction. Hence, if an honest client commits $tx$, no conflicting transactions can be committed.


Termination can be guaranteed when clients are honest and do not submit conflicting transactions. Honest clients will submit $tx$ to all honest blockchains, and the $tx$ will be eventually included in each of them. Given the termination of the ledger consensus running on local blockchains, the check function will return true eventually.

\begin{algorithm}[H]
\caption{Protocol \tblite$(\plocal_k)$ }
\label{alg:tblite}
\begin{algorithmic}[1]
    \Init
        \State $m \gets$ number of participating chains
    \EndInit
    
\Function{{\sf submit}}{$tx$}
    \For{$dst = 1, \cdots, m$}
        \State invoke $\plocal_k.\mathsf{submit}(tx)$\label{lite:submit}
    \EndFor
\EndFunction

\Function{{\sf check}}{$tx$}
    \State $cnt\gets 0$
    \For{$k = 1, \cdots, m$} \Comment{These $m$ queries are executed concurrently }
        \If{$\plocal_k.${\sf check}$(tx)$ returns true}
            \State $cnt \gets cnt + 1$
        \EndIf
    \EndFor
    \If{$cnt\ge \lfloor 2m/3\rfloor + 1$ and {\sf valid}$(tx)$} \label{lite:check}
        \State return true
    \EndIf
    \State return false
\EndFunction

\Function{{\sf valid}}{$tx$}
    \State $(in, out)\gets tx$
    \State $valid\gets$ true
    \For{$input \in in$}
        \If{{\sf check}$(input.from)$ returns false}
            \State $valid \gets$ false
        \EndIf
    \EndFor
    \State return $valid$
\EndFunction

\end{algorithmic}
\end{algorithm}




\section{Implementation}
\label{sec:impl}

We implement \tb in the {\sf Cosmos} ecosystem~\cite{kwon2019cosmos}, which is a decentralized network of parallel and interoperable blockchains, each powered by BFT consensus protocols like Tendermint~\cite{buchman2016tendermint}, where the CCC is enabled by an inter-blockchain communication (IBC) protocol~\cite{goes2020interblockchain}. In this section, we first give a brief overview of the {\sf Cosmos} ecosystem and then highlight the key challenges to implement \tbnosp.

\subsection{{\sf Cosmos} overview}

\noindent{\bf {\sf Cosmos} SDK.} {\sf Cosmos}-SDK~\cite{cosmos} provides tools for building permissioned or proof-of-stake (PoS) blockchains.
{\sf Cosmos}-SDK allows developers to easily create custom programmable and interoperable blockchain applications within the {\sf Cosmos} network without having to recreate common blockchain functionality.
{\sf Cosmos}-SDK has several pre-built modules to serve different functionalities such as defining transactions, handling application state and the state transition logic, etc. The most important modules related to \tb are the CosmWasm module~\cite{cosmwasm} and the IBC module~\cite{ibc} (see below).

\begin{figure*}
    \centering
    \includegraphics[width=0.7\textwidth]{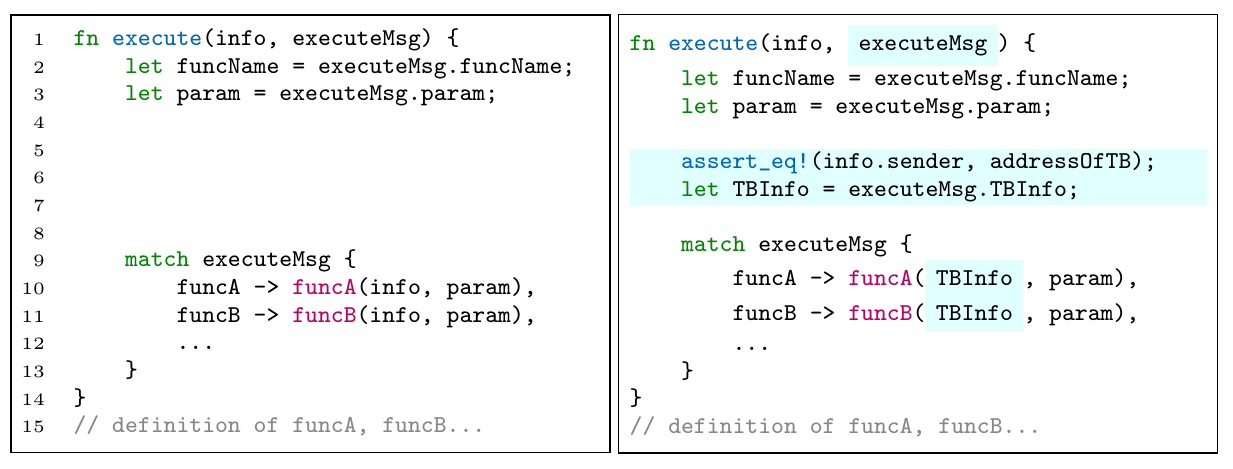}
     \caption{Left: Handler of CosmWasm contract. Right: Handler of \appx contract with \tb proxy. }
      \label{fig:entry}
\end{figure*}

\noindent{\bf CosmWasm.} CosmWasm adds smart contract support to the {\sf Cosmos} chains, where Rust is currently the most used programming language for contracts. The basic function calls of a CosmWasm contract are executed through an entry point (or a handler) shown in Fig.~\ref{fig:entry} (Left), by processing two given parameters\footnote{We omit some semantic details for simplicity, check full descriptions in \cite{cosmwasm}.}. The $\mathtt{info}$ contains contract information about function executions such as the address of the transaction sender, while the $\mathtt{executeMsg}$ encapsulates the name and parameters of the target function, which will be processed by the handler using a pattern-matching statement.

\noindent{\bf IBC protocol.} IBC is an interoperability protocol for communicating arbitrary data between {\sf Cosmos} blockchains. The protocol consists of two distinct layers: the transport layer which provides the necessary infrastructure to establish secure connections and authenticate data packets between chains, and the application layer, which defines exactly how these data packets should be packaged and interpreted by the sending and receiving chains. In the transport layer, blockchains are not directly sending messages to each other over networking infrastructure, but rather are creating messages to be sent which are then physically relayed from one blockchain to another by monitoring ``relayer processes''. These relayers~\cite{relayer} continuously scan the state of chains that implement the IBC protocol and relay packets when these packets are present. This enables transaction execution on connected chains when outgoing packets relayed over have been committed. Relayers cannot modify IBC packets, as each IBC packet is verified using light-clients by the receiving chain before being committed.

\noindent{\bf {\sf Cosmos} ecosystem.} Currently, over twenty CosmWasm-enabled blockchains are connected in the {\sf Cosmos} ecosystem by the IBC protocol. Therefore, {\sf Cosmos} provides the ideal environment for us to build and deploy \tbnosp.

\subsection{\tb implementation.}

\begin{figure*}
    \centering
    \includegraphics[width=0.7\linewidth]{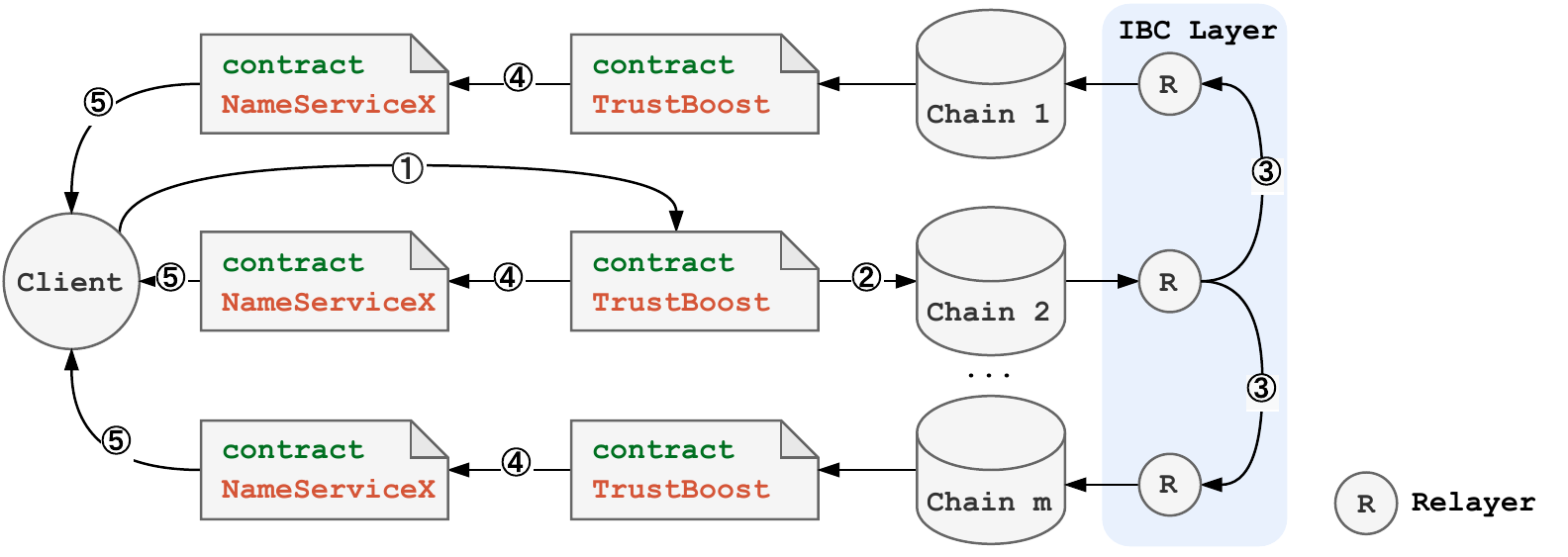}
    \caption{The flow chart of executing a \tb transaction.}
    \label{fig:flowchart}
\end{figure*}

We implement and deploy \tb as cross-chain smart contracts on $3f+1$ {\sf Cosmos} chains. It consists of two major parts, the \tb contract and cross-chain application contracts (denoted as {\sf AppX}). A complete flow chart is shown in Fig.~\ref{fig:flowchart}. To use application with \tbnosp, a client first calls the \tb contract to initiate a request to a specific application (\circled{1}), which is logged on the local blockchain (\circled{2}) and trigger a consensus protocol among all the blockchains (\circled{3}). Once the request is committed by \tbnosp, it calls the corresponding application contract to execute the request (\circled{4}). Clients can extract global states from each local contract (\circled{5}). In our implementation, the example application is the {\sf NameService} contract, where users can buy and transfer domain names. We also observe that the changes to turn a single-chain application contract into a cross-chain one are minor\footnote{See changes to upgrade a contract to have \tb support in \href{https://github.com/trustboost/cosmos-nameservice/commit/c59df344400dc915fd5907627e4fdf12a80eb325?diff=split}{github link}.}. 

Next, we discuss key challenges of implementing \tbnosp. 

\noindent{\bf Use PKI or not?} In the relatively recent blockchain era, partially synchronous BFT protocols have received renewed attention; performant and efficient BFT protocols have been constructed (e.g., SBFT~\cite{gueta2019sbft}, Tendermint~\cite{buchman2016tendermint} and HotStuff~\cite{yin2019hotstuff}). However, all these protocols require a public key infrastructure (PKI) or even a threshold signature scheme. 
Unfortunately, we found that in {\sf Cosmos}-SDK, once an IBC packet is verified using light-clients by the receiving chain, the signatures (i.e., the quorum certificate from the sending chain) are then removed and only the plaintext of the IBC message is passed to the receiving chain. Therefore, the IBC message verification in {\sf Cosmos} current lacks of transferability, that is a third chain won't be able to verify that an IBC message is indeed sent from its sending chain. 

Due to this limitation, we have to implement BFT protocols without using PKI and the state-of-the-art one is Information Theoretic HotStuff (IT-HS)~\cite{abraham2020information}. IT-HS has a quadratic message complexity in each view; and it is optimistically responsive: with a honest leader, parties decide as quickly as the network allows them to do so, without regard for the known upper bound on network delay. Although IT-HS has slightly higher round and message complexity than Tendermint/HotStuff, it avoids many expensive operations, such as signature verification and aggregation. In addition, IT-HS only requires constant persistent storage. Therefore, we believe IT-HS is a good candidate to be implemented as the consensus protocol for our \tb smart contract.

We also note that if we make a few minor changes to the {\sf Cosmos}-SDK (i.e., passing the signatures to the smart contract layer), then \tb can implement any BFT protocol. However, to make sure that \tb is directly deployable in the real world, we keep the {\sf Cosmos}-SDK untouched and stick to IT-Hotstuff in this paper.

\noindent{\bf Consensus as smart contracts.} To the best of our knowledge, this is the first work to build consensus protocols using smart contracts. The inherent limitations to  smart contract programming (e.g.,  single-threading) pose challenges to consensus protocol implementation. A prominent challenge is to ensure that every operation in the BFT protocol is properly triggered by some IBC message. Fortunately, this is the case in IT-HS (as it is optimistically responsive). The only exception is the timeout for the view change, which occurs when the leader is malicious or the network is poor, thus this is not in the optimistic path. Particularly, in IT-HS, a party will enter the view change phase if the leader of current view does not make any progress for a certain period of time (a pre-defined timeout value). However, this is not triggered by any IBC message. To address this issue, we set external bots/scripts to regularly ping the blockchains to trigger the timeout in time. Note that we make no trust assumptions on these bots (they can also be replaced by ``keepers'', a recent proposal from   Chainlink~\cite{keeper}).

Another limitation of the IBC protocol is that a blockchain can not send IBC messages to itself. However, in many BFT protocols including IT-HS, there are operations triggered by self-delivered messages. Therefore, we have to pay special attention to the self-delivered messages when implementing IT-HS in \tbnosp. In addition, {\sf Cosmos}-SDK allows the smart contract to send IBC messages only when a function returns; once that is complete, the function call is over and no more operations can be conducted. To deal with these constraints in {\sf Cosmos}, we use a {\it message queue} in each function, queue all the IBC messages that need to be sent, and send them all when the function returns. For example, in IT-HS, when a node receives a {\sf propose} message from the current leader, it will send a {\sf echo} message to all the other nodes; and when a node receives $2f+1$ {\sf echo} messages, it will send a {\sf key1} message to all the other nodes. Extra caution is warranted when writing the {\sf send\_proposal} function in \tbnosp: in  the {\sf send\_proposal} function, the current leader blockchain would like to broadcast the {\sf propose} message, but can only do this when the function returns. So, we need to first put the {\sf propose} message into the message queue. Considering that the leader blockchain should send an {\sf echo} message when the {\sf propose} message is self-delivered, we also put the {\sf echo} message into the message queue. And because the {\sf echo} message will also be self-delivered, we increase the counter for {\sf echo} messages by one. Since there are not enough {\sf echo} messages yet, the next step (sending the {\sf key1} message) will not be triggered, the {\sf send\_proposal} function can finally return and the leader blockchain can send the IBC messages in the message queue ({\sf propose} and {\sf echo}). Several  such subtle aspects abound in the \tb implementation,  making  the design and implementation a  challenging and rewarding endeavor; as such, we believe  how to implement complex communication/distributed protocols using smart contracts is  of independent interest.

\noindent{\bf \tb as a proxy for application contracts.} In order to boost the security promises, any single-chain application contract \app can be equipped with a \tb proxy to become a cross-chain application contract \appx without touching functional codes. Specifically, a \tb proxy will issue a function call to \appx contract after committing a client's request. To support this, \appx contract only modifies a few lines (highlighted in Fig.~\ref{fig:entry} (Right)) in the handler function of \app contract to (1) check the function call is initiated by a certified \tb contract; and (2) add contract information \texttt{info} of \app contract as an extra parameter of \appx contract (stored in \texttt{executeMsg.TBInfo}) to reproduce single-chain executions.

\noindent{\bf Contract state.} Though an application contract \app and its cross-chain variant \appx provide the same functionality, they have to maintain isolated states. Regardless of the states owned by an \app contract already existing on any single blockchain, the deployment of a new \appx contract initializes all related states independently from scratch, which are secured by more stringent security rules.





\section{Evaluation}
\label{sec:eva}

Our experimental evaluation answers the following questions:
\begin{itemize}
    \item What is overhead of \tb in terms of gas usage and confirmation latency? This is the price paid for security.
    \item How well does \tb scale when the number of chains increases? 
    \item How does \tb perform under Byzantine attacks?
\end{itemize}

\noindent {\bf Testbed setup.} We deploy \tb on an AWS m5.4xlarge instance with 16 vCPU and 64 GB memory. There are three steps in the setup phase: 1) start $m=3f+1$ {\sf Cosmos} chain instances in our local testnet; 2) deploy \tb and {\sf NameService} contracts on each of the $m$ chains; 3) connect each pair of the $m$ chains with an IBC channel. In our experiments, the block rate of each {\sf Cosmos} chain is set to be about one block per second.


\noindent {\bf Performance.} In this experiment, we evaluate the performance of \tb with $m = 4,7,10$. We measure the number of IBC messages, the total gas usage and the confirmation latency per request. For the confirmation latency, we record the duration between the submission and the execution of the request. We repeat the  experiment on the single-chain {\sf NameService} itself without using \tb (i.e., $m=1$) for comparison (results in Table~\ref{tab:perf}).

\begin{table}
    \centering
    \begin{tabular}{|c|c|c|c|c|}
        \hline
        m & 1 & 4 & 7 & 10 \\
        \hline
        \hline
        \# IBC  & 0 & 102 & 348 & 738\\
        \hline
        Gas usage & 202K & 74M & 261M & 586M \\
        \hline
        Latency & 2.5s &  67.2s & 105.0s & 138.2s \\
        \hline
    \end{tabular}
    
    \caption{Performance of \tb with different number of chains.\label{tab:perf}}
\end{table}
\begin{table}
    \centering
    \begin{tabular}{|c|c|c|c|c|}
        \hline
        Attacks & I & II & III & IV \\
        \hline
        \hline
        Latency & 137.2s & 138.6s  & 66.8s & 66.0s \\
        \hline
    \end{tabular}
    \caption{Performance of \tb under different attacks.\label{tab:sec}}
\end{table}



From Table~\ref{tab:perf}, we see that the number of IBC messages and the gas usage scale quadratically in the number of chains. The overhead in gas comes from two parts: 1) the communications and computations in the \tb contract cost gas; 2) the {\sf NameService} contract needs to be executed on all $m$ chains. Note that for fixed $m$, the former gas usage is a constant, independent of the application contracts. And in our experiments, the {\sf NameService} contract uses very little gas, so the overhead caused by \tb is dominating. Further, by batching the requests, this overhead can be amortized. Also note that the performance of \tb is also limited by IBC efficiency: sending one single IBC message to transfer tokens cross chains would need 2s and 350K gas. Hence, making the transmission and execution of IBC messages more efficient could greatly improve the performance of \tbnosp.
Moreover, 600M gas only costs \$2-\$10, for example  based on the gas price and the exchange rate on the {\sf Osmosis} chain, at the time of writing (April 2023).
The latency is independent of the application contract. However, we can see that it scales almost linearly when $m$ increases, and it is pretty acceptable compared with other high security chains, such as {\sf Ethereum}. 


\noindent {\bf Security.} In this experiment, \tb is evaluated under active attacks, specifically within a four-chain scenario where one blockchain is compromised by an attacker and may behave arbitrarily. The malicious behaviors of the compromised blockchain are triggered by external calls. Just like many other BFT protocols, IT-HS also has a view-change sub-protocol to ensure liveness: when no progress is made in one view, all nodes will enter the next view by broadcasting {\sf abort} messages; each view is assigned with a predefined primary node. We test the following attacks.  
\begin{itemize}[leftmargin=*]
    \item {\bf Attack I - Primary blockchain crashes.} In this attack, the primary blockchain of the first view crashes at the beginning, therefore progress can only be made in the second view. 
    \item {\bf Attack II- Primary blockchain equivocates.} In this attack, the primary blockchain of the first view sends different proposals to different non-primary blockchains in the network. 
    \item {\bf Attack III - Non-primary blockchain equivocates.} In this attack, a non-primary blockchain of the first view sends different votes to different blockchains in the network. 
    \item {\bf Attack IV - Non-primary blockchain aborts.} In this attack, a non-primary blockchain of the first view keeps sending {\sf abort} messages to enter the view change phase.
\end{itemize}

We measure the confirmation latency under these attacks. The results are shown in Table~\ref{tab:sec}. We can see that \tb still works under these attacks: the confirmation latency doubles under the first two attacks as it takes two views to terminate; and attacks III and IV hardly have any impact on the latency.

\section{Discussion}
\label{sec:dis}

\noindent {\bf Unequal weights.} Just like many BFT-style PoS protocols~\cite{gilad2017algorand}, blockchains in \tb can also have different weights, i.e., the voting powers in the BFT protocol. For example, if all constituent chains are PoS chains, we can set the weight of a chain to be proportional to the total market cap of its native token. 
This strategy is also adopted in recent research \cite{tas2023interchain}, which interprets the stake required for an adversary to compromise safety as a measure of economic security. This enables the comparison of security levels among various PoS chains, aiming to maximize the expense required to attack the \tb ledger. Another example is that we can set the weight of all chains except one strong chain to be zero so that \tb can directly borrow trust from the strong chain.  How to dynamically adjust the weights is also an interesting question. Moreover, today many blockchains are heavily intertwined economically (e.g., {\sf Osmosis} and {\sf Axelar Network} in {\sf Cosmos}), so the idea of asymmetric trust~\cite{cachin2021asymmetric} can also be brought to \tbnosp. We defer these topics to the future work.

\noindent {\bf Share security via checkpointing.}  An important application of \tb is that we can use it to checkpoint the $m$ constituent chains or other weak chains. The validators of each chain just need to regularly submit block hashes and signatures as checkpoints to the \tb ledger, and the finality rule of each chain will be altered to respect the checkpoints. In this way, each chain will also have a slightly slower finality rule - confirming the chain up to the latest checkpoint, which has the same latency and security level as \tbnosp. For high value transactions on a constituent chain, the users can apply the slow finality rule to enjoy stronger security guarantees. In the context of {\sf Cosmos}, with this approach each constituent {\sf Cosmos} chain in \tb will have slashable safety and much shorter withdrawal delays as long as at most $1/3$ of the chains are faulty, following the results shown in~\cite{tas2022bitcoin}. 

\noindent {\bf Cross chain applications.} We note that an important application of cross chain bridges today is cross chain token transfer. However, it has a fundamental security limit, where the attacker transfers some tokens on chain 1 to chain 2 and then reverts the state on chain 1 (e.g., by doing 51\% attack) to get his tokens back. We point out that this issue can be alleviated by \tb using the checkpointing idea discussed above. Particularly, when a user wants to move 100 token ABC from chain 1 to get 100 token XYZ on chain 2, first the 100 token ABC will be locked on chain 1 and then token XYZ will be sent to the user only when the lock transaction on chain 1 is confirmed by the slow finality rule (i.e., in the checkpointed chain). In this way, the state of the token ABC contract on chain 1 is implicitly upgraded into the global state of the \tb ledger, which has stronger security guarantees. 

\noindent {\bf Heterogeneous chains.} Although our \tb implementation is in the {\sf Cosmos} ecosystem, the idea can be extended to a heterogeneous blockchain network. We just need all the heterogeneous chains to be programmable and interoperable. Different blockchains may have different virtual machines, therefore the \tb smart contract will need to be written in multiple programming languages. On the other hand, there are quite a few ongoing projects using zkSNARKs to build trustless and efficient cross chain bridges~\cite{snark,xie2022zkbridge}. We believe \tb will have broader applications in the near future when the heterogeneous blockchain network becomes mature.

\section{Acknowledgements}

We thank Dion Hiananto and Luhao Wang for their help with the implementation and experiments. We thank Jack Zampolin and a few other {\sf Cosmos} core developers for valuable suggestions on this project. This research is supported in part by the US National Science Foundation under grants CCF-1705007 and CNS-1718270, the US Army Research Office under grant W911NF1810332 and W911NF2310147 and a gift from XinFin Private Limited.

\bibliographystyle{plain}
\bibliography{ref}

 \appendix
 \section{The Limitation of Ledger Combiner}
\label{sec:ledger-combiner}
\begin{figure}
    \centering
    \includegraphics{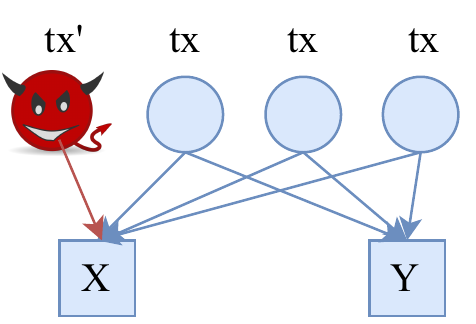}
    \caption{An example with $m=4$ objects and $f=1$ Byzantine object. All honest objects receive and commit the same transaction $tx$; only clients in $X$ see the conflicting transaction $tx'$ committed on the Byzantine object.}
    \label{fig:combined_ledger}
\end{figure}

The previous work on robust ledger combiner \cite{fitzi2020ledger} proposes that in passive mode, when $m\ge 2f + 1$, a combined ledger with relative settlement can be constructed. This relative settlement definition asserts that if a transaction is relatively settled, it will remain reliably settled in the ledger from that point forward, and no conflicting transactions will appear in the ledger unconsciously. However, this definition is notably less robust than our definition (see Definition \ref{def:abc}) of ABC. Specifically, ABC includes the assurance of honest termination, meaning that if every honest object receives the same transaction (and no conflicting ones), that transaction will be committed by all honest processes. Unfortunately, relative settlement falls short of guaranteeing this essential property.

Consider the example illustrated in Figure~\ref{fig:combined_ledger}, where $m=4$ and $f=1$. Here, three honest objects receive the identical transaction $tx$, and each commits $tx$. Processes in groups $X$ and $Y$ both observe this. However, suppose a Byzantine object commits a conflicting transaction $tx'$ but only reveals it to group $X$. In our \tblite protocol, both $X$ and $Y$ will commit $tx$, as honest objects received the same transaction and no conflicting transactions. In contrast, under the ledger combiner model of \cite{fitzi2020ledger}, only group $Y$ will commit $tx$, while group $X$ will not commit $tx$ because they observed a conflicting transaction $tx'$. This divergence violates the honest termination property.

Furthermore, Theorem \ref{thm:passive} in our work presents a profound implication: ABC is unachievable when $m \le 3f$. Since the model in \cite{fitzi2020ledger} assumes $m \ge 2f + 1$, it follows that the protocol they present cannot successfully solve ABC according to our theoretical lower bound.

\end{document}